\documentstyle[prl,aps,floats]{revtex}
\begin{document}
\twocolumn[
\begin{centering}
{\large \bf Evolution and global collapse of trapped Bose
condensates under\\
variations of the scattering length}
\vspace{1.5ex}\\
Yu. Kagan$^{1}$, E.L. Surkov$^{1}$,
and G.V. Shlyapnikov$^{1,2}$\vspace{1ex}\\

{\em (1)} {\it Russian Research Center Kurchatov Institute,
Kurchatov Square, 123182 Moscow, Russia}\\
{\em (2)} {\it FOM Institute for Atomic and Molecular Physics,
Kruislaan 407, 1098 SJ Amsterdam, The Netherlands}\vspace{0.5ex}\\

\end{centering}
\begin{quote}

{\small We develop the idea of selectively manipulating the
condensate in a
trapped Bose-condensed gas, without perturbing the thermal cloud.
The idea is based on the possibility to modify the mean field
interaction
between atoms (scattering length) by nearly
resonant incident light or by spatially uniform change of the
trapping
magnetic field.
For the gas in the Thomas-Fermi regime we find analytical scaling
solutions
for the condensate wavefunction evolving under arbitrary variations
of the
scattering length $a$.
The change of $a$ from positive to negative induces a global
collapse of the condensate, and the final stages of the collapse
will be
governed by intrinsic decay processes.

}
\end{quote}
]

\vspace{4mm}

\narrowtext

The discovery of Bose-Einstein condensation (BEC) in trapped clouds
of
ultra-cold alkali atoms \cite{Cor95,Hul95,Ket95} opened unique
possibilities
to investigate collective many-body effects in dilute gases.
Isolation of a trapped gas from the environment, provided by a
wall-less magnetic confinement, makes this object attractive for
studying fundamental problems of the physics of many-body quantum
systems, such as relaxation and the loss of coherence in the
evolving
macroscopic quantum state (condensate).
In ongoing experiments \cite{jila,And96,Mew96,mit,Jin96} the
condensate
is set into motion
(for example, undergoes oscillations) by varying the confining
field.
Scaling theory of coherent evolution (without damping and
relaxation) of a
Bose condensate in harmonic traps under arbitrary frequency
variations has been developed in \cite{Kagan96,Castin}.
It is important, however, that variations of the confining potential
also cause the evolution of the thermal component of a trapped gas.
Moreover, in the hydrodynamic regime the thermal cloud in
many aspects evolves similarly to the condensate \cite{Kagan96}.
For example, the asymmetry of free expansion and eigenfrequencies of
small oscillations are almost the same.

In this paper we develop the idea of selectively manipulating the
condensate, without perturbing the thermal cloud.
This is especially important for studying the interaction between
the
condensate and the thermal component.
The idea is based on the possibility to modify the mean field
interaction between atoms.
At low temperatures the latter is proportional to the scattering
length $a$ which, as found in \cite{Fed96}, can be changed
under the influence of red-detuned nearly resonant light.
Another option to modify the scattering length relies on the
magnetic
field dependence of $a$, predicted in \cite{Verhaar}, and assumes
spatially uniform variations of the trapping field, without changing
the trap frequencies.
Since the shape of the condensate wavefunction is predetermined by
the
interaction between particles, the change of $a$ will cause
the evolution of the condensate at constant frequencies of the
confining
potential.
Remarkably, at temperatures $T\gg n\tilde U$ ($n$ is the gas
density,
$\tilde U=4\pi\hbar^2a/m$, and $m$ the atom mass) only the
condensate
evolution will be pronounced, which resembles the picture of the
fourth
sound in superfluid helium.
In this temperature range the shape of the thermal cloud is mainly
determined by the trapping potential and temperature.
The perturbation of the thermal component will be small compared to
that of the condensate, at least as $n\tilde U/T$.

The change of $a$ in the field of nearly resonant red-detuned light
is provided by virtual radiative transitions of a
pair of atoms to a bound electronically excited molecular state.
Since the resonance dipole interaction between atoms in the excited
state is
much stronger than the interaction in the ground state, already for
moderate light intensities the effective interatomic interaction
and, hence, the scattering length can be significantly changed.
It is even possible to switch the sign of $a$ \cite{Fed96}.

The magnetic field dependence of the scattering length, found in
\cite{Verhaar}, has a resonance structure.
Therefore, spatially uniform variations of the field can also change
$a$
substantially and reverse its sign.

Below we find analytical scaling solutions for the condensate
wavefunction evolving under arbitrary variations of the scattering
length.
We consider the most interesting situation, where the mean field
interaction
between atoms greatly exceeds the level spacing in the trap
(Thomas-Fermi
regime).
The case of switching $a$ from positive to negative and inducing a
global
collapse of the condensate is specially analyzed.

Let us consider the evolution of a condensate in a harmonic
potential $V({\bf r})=m\sum_i\omega_i^2r_i^2/2$, with constant
frequencies
$\omega_i$, induced by variations of the mean field interaction
between
particles.
The equation for the condensate wavefunction $\Psi_0({\bf r},t)$,
with above-condensate particles neglected, can be represented in the
form
\begin{equation}  \label{Schred2}
i\hbar\frac{\partial\Psi_0}{\partial t}\!=\!-\frac{\hbar^2}{2m}
\Delta\Psi_0\!+\!
\frac{m}{2}\sum_i\omega_i^2r_i^2\Psi_0\!+\!\tilde
U(t)|\Psi_{0}|^2\Psi_0.
\end{equation}
Here $\tilde U(t)=4\pi\hbar^2a(t)/m$, and $a(t)$ is the
time-dependent
scattering length.
Along the lines of the general method developed in \cite{Kagan96},
we
introduce $d$ scaling parameters $b_i(t)$, where $d$ is the
dimension of the
system.
Turning to new coordinate and time variables
$\rho_i\!=\!r_i/b_i(t)$,
$\tau(t)\!\!=\!\!\int^t\!dt'\gamma(t')/{\cal V}(t')$,
with $\gamma(t)=\tilde U(t)/\tilde U_0$ and
$\tilde U_0$ the initial value of $\tilde U(t)$, we search for the
solution
in the form
\begin{equation}              \label{nop}
\Psi_0({\bf r},t)={\cal V}^{-1/2}(t)\chi_0(\mbox{\boldmath
$\rho$},\tau(t))
\exp{(i\Phi({\bf r},t))},
\end{equation}
where the dimensionless volume ${\cal V}(t)\!=\!\prod_ib_i(t)$.
Substituting Eq.(\ref{nop}) into Eq.(\ref{Schred2}) we require the
cancellation of the $\nabla_{\rho}\chi_0$ terms. This leads to the
relation for the phase:
\begin{equation}            \label{phase}
\Phi({\bf r},t)\!=\!(m/2\hbar)\!\sum_i\!r_i^2[\dot b_i(t)/b_i(t)].
\end{equation}
The scaling parameters $b_i(t)$ will be chosen such that they are
governed by the equations
\begin{equation}             \label{bgen}
\ddot b_i+\omega_i^2b_i=\omega_{i}^2\gamma(t)/b_i{\cal V}(t),
\end{equation}
with initial conditions $b_i(0)=1$, $\dot b_i(0)=0$.
Then, we arrive at the equation of motion
\begin{equation}              \label{chigen}
\!\!i\hbar\frac{\partial\chi_0}{\partial\tau}\!=\!\!-\frac{\hbar^2}{2m}
\!\!\sum_i\!\!\frac{{\cal V}(t)}{\gamma(\!t) b_{\!i}^2\!(\!t)}
\frac{\partial^2\!\chi_0}{\partial\rho_{i}^2}\!+\!
\frac{m}{2}\!\!\sum_i\!\omega_i^2\rho_i^2\!\chi_{\!0}\!\!+\!\!\tilde
U_{\!0}
|\chi_{\!0}\!|^2\!\chi_{\!0}.\!\!\!
\end{equation}

In the case of initially repulsive ($a(0)>0$) and strong interaction
between particles (Thomas-Fermi regime) the initial chemical
potential $\mu_0=n_0\tilde U_0\gg \hbar\omega_{i}$ ($n_0$ is the
maximum
density in the initial static condensate), and the kinetic energy
term
in Eq.(\ref{chigen}) is initially small compared to the non-linear
interaction term .
In the course of evolution the ratio of the kinetic to interaction
term
scales as
$\varepsilon(t)=\sum_i(\hbar\omega_{i}/\mu_0b_i(t))^2{\cal
V}(t)/\gamma(t)$.
Assuming that the condition $\varepsilon (t)\ll 1$ is satisfied at
any $t$,
the kinetic energy term can be omitted.
In the case of a fast change of the interparticle interaction this,
in
particular, requires the inequality $|\gamma|\!\gg\!
(\hbar\omega_i/\mu_0)^2$ at $t$ close to $0$.
Then in the variables $\rho_i$, $\tau$ Eq.(\ref{chigen}) is reduced
to
that for the static case, with initial interaction between
particles.
The solution has the well known form (see \cite{Silvera,Huse}):
\begin{equation}             \label{pargen}
\chi_0(\mbox{\boldmath $\rho$},\tau(t))\!=
\!\frac{1}{\tilde U_0^{1/2}}\!\left(\!
\mu_0\!-\!\frac{m}{2}\!\sum_i\!\omega_i^2\rho_i^2\!\right)^{1/2}
\!\!\!\!\!\!\!\!\!\exp{\!\left(\!\frac{-i\mu_0\tau(t)}{\hbar}\!\right)
}\!
\end{equation}
in the spatial region where the argument of the square root is
positive
and zero otherwise.
Thus, the condensate evolution under arbitrary variations of the
scattering length is described by a universal scaling solution for
$\Psi_0({\bf r},t)$, following from Eqs.~(\ref{nop}) and
(\ref{pargen}).
Actually, the problem is reduced to the solution of classical
equations
(\ref{bgen}) for the scaling parameters $b_i(t)$ (cf.
\cite{Kagan96}).

We should emphasize that the results for $d=2$, obtained below, can
be
applied equally well to the radial evolution of $\Psi_0$ in very
long
samples where the axial frequency is much smaller than the radial
one.
This follows from the fact that then to a first approximation the
dependence
of $\Psi_0$ on the axial coordinate can be omitted.

In the limiting case of adiabatically slow variations of the
scattering
length, that is on a time scale $\tau_0\gg \omega_i^{-1}$, the
solution of
Eqs.(\ref{bgen}) leads to simple relations
\begin{equation}      \label{bad}
b_i(t)=[\gamma(t)]^{1/(2+d)}\equiv b(t);\,\,\,\,\,\,\,\,
{\cal V}(t)=[\gamma(t)]^{d/(2+d)}.
\end{equation}
Assuming that at $t\!\gg\!\tau_0$ the scattering length acquires a
constant
value $a_1$ ($\tilde U(t)\!=\!\tilde U_1$, $\gamma(t)\!=\!\tilde
U_1/\tilde
U_0\!\equiv\!\gamma_1$), for the phase in Eq.(\ref{pargen}) we
obtain
$\mu_0\tau(t)\!=\!\mu_1t$, where the quantity $\mu_1\!=\!\mu_0
\gamma_1^{2/(2+d)}\!\!$
is equal to the chemical potential of the gas with the interparticle
interaction $\tilde U_1$.
Then, from Eqs.~(\ref{pargen}) and (\ref{nop}) one can see that the
initial condensate is adiabatically transformed to the
static condensate with the wavefunction
corresponding to the interaction $\tilde U_1$.

For an abrupt change of the interparticle interaction from $\tilde
U_0$
to $\tilde U_1$ scaling equations (\ref{bgen}) take the form
\begin{equation}               \label{bfast}
\ddot b_i+\omega_i^2b_i=\omega_i^2\gamma_1/b_i{\cal V}(t).
\end{equation}
In fact, for $\gamma_1=\tilde U_1/\tilde U_0 >0$ these equations are
similar to those
for the scaling parameters of the condensate evolution after an
abrupt
change
of the trapping potential, with initial frequencies being
$\omega_i\sqrt{\gamma_1}$ and final frequencies $\omega_i$ (see
\cite{Kagan96}).
The solution of Eqs.(\ref{bfast}) gives oscillating functions
$b_i(t)$
which ensure undamped oscillations of the condensate wavefunction.
For $\gamma_1$ close to $1$ the frequency spectrum of the condensate
oscillations coincides with the set of eigenfrequencies of small
shape
oscillations of the initial condensate, found for cylindrically
symmetric traps in the JILA \cite{jila} and MIT \cite{mit}
experiments
and calculated in \cite{Stringari,Burnett,Castin,Kagan96}.
It is important that an abrupt change of the scattering length is
equivalent
to equal relative change of all frequencies.
This reduces the possibility of stochastization of the condensate
evolution in an anisotropic harmonic potential, compared to the case
of
independent change of the frequencies.

The structure of the condensate oscillations caused by the change of
the scattering length can be easily analyzed in the case of
isotripic
trapping configuration.
For $d=2$ Eq.(\ref{bfast}) has analytical solution
\begin{equation}            \label{b2}
b_i^2(t)=b^2(t)=[1+\gamma_1+(1-\gamma_1)\cos{2\omega t}]/2.
\end{equation}
Once the scattering length is decreased ($\gamma_1<1$) there will be
"compression oscillations": Compression of the condensate will
be followed by its expansion to the initial shape.
An increase of $a$ ($\gamma_1>1$) induces "expansion oscillations",
i.e.
the condensate is expanding and then compressing to the initial
shape.
In both cases the scaling parameter $b$ varies from $1$ to
$\sqrt{\gamma_1}$.
In the $3$-d case the condensate oscillations are anharmonic, with a
characteristic period somewhat smaller than that for $d=2$.
The parameter $b$ is varying from $1$ to $\sqrt{2\gamma_1/3}$.

The dynamics of the system, induced by the change of the
interparticle
interaction, is drastically different from that caused by variations
of
the trap frequencies.
In the former case the condensate oscillates, whereas the thermal
component is only weakly perturbed and practically remains static.
In the latter case the thermal cloud is fully involved in the
oscillatory
evolution process.
Preferential evolution of the condensate under the change of the
scattering
length resembles the fourth sound in superfluid helium, where the
superfluid part of the density oscillates on the background of a
static
normal component.
As well as in helium, the analysis of dynamic properties of a
trapped gas in
these conditions can be an effective method of identifying and
studying BEC.

The change of the interaction between particles can lead to an
interesting phenomenon, a global collapse of the condensate as a
whole.
Let us assume that initially the scattering length $a>0$, and the
system is in a stationary state.
The sign of $a$ can be switched to negative in the field of nearly
resonant
red-detuned light \cite{Fed96} or by
spatially uniform variations of the confining magnetic field (see
\cite{Verhaar}).
Then, for an abrupt change of $a$ the solution of scaling equations
(\ref{bfast}) immediately
leads to a self-compression (collapse) of the condensate.
In the $d=2$ isotropic case this directly follows from the exact
solution (\ref{b2}): For $\gamma_1<0$ the scaling parameter $b$
decreases and reaches zero at $t=t_{*}$, where
\begin{equation}             \label {tstar}
t_{*}=\omega^{-1}\arcsin{(1/\sqrt{1+|\gamma_1|})}.
\end{equation}
In the $3$-d isotropic case $t_{*}$ is close to that from
Eq.(\ref{tstar}).
In the vicinity of $t_{*}$, where $\omega\Delta t\!\!\ll\!\!{\rm
min}
[|\gamma_1|^{\!{1\!/\!d}}\!\!,\!|\gamma_1|^{\!{-1\!/\!2}}]$ ($\Delta
t\!=\!t_{*}\!-\!t$), the solution of Eq.(\ref{bfast}) takes the form
\begin{equation}             \label{bstar}
b^2(t)=[(\sqrt{|\gamma_1|}\omega\Delta t)(2+d)/\sqrt{2d}]^{4/(2+d)}.
\end{equation}

In both cases the compression rate increases with decreasing $\Delta
t$:
The quantity $\dot b(t)/b(t)=2/[(d+2)\Delta t]$, and the
characteristic
velocity of the condensate boundary, $v\propto [\Delta
t]^{-d/(d+2)}$.
In the course of collapse the small parameter $\varepsilon(t)$
remains
constant ($d=2$) or even decreases ($d=3$), which justifies the
neglect of the kinetic energy term in Eq.(\ref{chigen}).
The kinetic energy of the system is determined by the behavior of
the
phase $\Phi({\bf r})$ (\ref{phase}) which drastically increases with
decreasing $\Delta t$.
A strong rise of the kinetic energy in the course of collapse is
compensated by decreasing potential energy ($a<0$), which ensures
the conservation of the total energy.

The increase of density in the collapsing condensate enhances
intrinsic inelastic processes, such as three-body recombination and
spin-dipole relaxation in binary collisions.
This leads to particle losses, since fast atoms and
molecules produced in the inelastic processes escape from the trap.
We will confine ourselves to the analysis of the influence of
inelastic processes on the dynamics of collapse in the isotropic
$d=2$
and $d=3$ cases.
The atom loss rate is determined by the relation
\begin{equation}             \label{aloss}
\dot N=-\alpha_s\int d^dr|\Psi_0({\bf r},t)|^4-\alpha_r
\int d^dr|\Psi_0({\bf r},t)|^6,
\end{equation}
where $\alpha_s$ and $\alpha_r$ are the rate constants of spin
relaxation and three-body recombination, respectively.
With increasing density in the course of collapse, the ratio of the
recombination to relaxation rate increases.
In ultra-cold alkali atom gases $\alpha_s$ is in the range
$10^{-14}-10^{-16}$ cm$^3$/s and $\alpha_r\sim 10^{-28}$
cm$^6$/s.
Accordingly, considering initial condensate densities $n_0\sim
10^{13}-10^{14}$ cm$^{-3}$ (currently achieved in Rb and Na BEC
experiments \cite{jila,And96,Mew96,mit,Jin96}),
we see that already for moderate compression of the condensate
three-body recombination dominates over spin relaxation.
Therefore, the process of spin relaxation is omitted in the
further analysis.
For the same reason, in the case of light-induced change of the
scattering length  we omit the process of
photoassociation in pair collisions.

The loss of particles from the condensate influences the evolution
of the condensate wavefunction.
The character of the evolution depends on the density of the
collapsing condensate, $n(t)$, and is predetermined by the relation
between the characteristic time of three-body recombination,
$\tau_r\sim (\alpha_r n^2)^{-1}$, and the correlation time
$\tau_c\sim\hbar/n|\tilde U_1|$. The latter is responsible for
establishing the shape of the condensate wavefunction corresponding
to the instantaneous value of the number of particles $N(t)$.
There will be two different stages of the evolution.
The role of three-body recombination becomes especially important at
densities
\begin{equation}             \label{nstar}
n\agt n_{*}\equiv |\tilde U_1|/\hbar\alpha_r\gg n_0,
\end{equation}
where the compression is very strong and $\tau_r\alt\tau_c$.
Then already the shape of $|\Psi_0({\bf r},t)|$ is determined by the
recombination losses (see below).
With realistic numbers $|a_1|\sim 10$ \AA, and $\alpha_r\sim
10^{-28}$
cm$^{6}$/s, we see that the density $n_{*}$ will be in the range
$10^{17}-10^{18}$ cm$^{-3}$ and still satisfies the criterion of
weakly
interacting gas, $n_{*}|a|^3\ll 1$.

In the density range $n\ll n_{*}$ the recombination time $\tau_r\gg
\tau_c$ and the loss of particles occurs in a quasistationary
regime.
For $\Psi_0({\bf r},t)$ one can use the
scaling solution following from Eqs.~(\ref{nop}), (\ref{pargen}),
with instantaneous values of $N$ and the chemical potential $\mu$.
Since $\mu\sim N^{2/(2+d)}$, for the time dependence of the maximum
condensate density in the quasistationary stage we obtain
\begin{equation}            \label{nmax}
n(t)=n_0[b(t)]^{-d}[N(t)/N_0]^{2/(2+d)},
\end{equation}
where $N_0$ is the initial number of condensate particles.

Below we assume that $|\gamma_1|\alt 1$.
Then the characteristic time of collapse, $t_{*}\sim \omega^{-1}$.
Since for realistic parameters of the system the initial
recombination time $\tau_{r0}\gg \omega^{-1}$, a major part of the
compression occurs without particle losses.
This is the case even at $t$ rather close to $t_{*}$, the
condensate density being determined by Eq.(\ref{nmax}) with the
scaling parameter $b$ from Eq.(\ref{bstar}) and $N\approx N_0$:
\begin{equation}           \label{nmax1}
n(t)\approx n_0(\sqrt{|\gamma_1|}\omega\Delta t)^{-2d/(2+d)}.
\end{equation}
Eq.(\ref{nmax1}) is no longer valid when the number of lost
particles
becomes comparable with $N_0$.
The characteristic time $\Delta t_L$ and the density $n_L$ at which
this
happens can be found from Eq.(\ref{aloss}), with $\Psi_0$ determined
by
Eqs.~(\ref{nop}), (\ref{chigen}) and the scaling parameter $b(t)$
given by Eq.(\ref{bstar}):
$$\!\dot N\!\!=\!\!-\frac{\alpha_r}{\tilde
U_0^3}\!\!\int\!\!\!\frac{d^d\!\rho}
{b^{2d}(t)}(\mu_0\!-m\omega^2\!\rho^2\!/2)^3\!\approx\!\!-
\frac{N_0}{\tau_{\!r0}(\!\sqrt{|\gamma_1\!|}\omega\Delta
t)^{4d/\!(\!2+d)}}.$$
Comparing the number of lost atoms with $N_0$ we obtain
\begin{eqnarray}
\sqrt{|\gamma_1|}\omega\Delta t_L\approx (\sqrt{|\gamma_1|}
\omega\tau_{r0})^{-(2+d)/(3d-2)};    \label{t1} \\
n_L\approx n_0(\sqrt{|\gamma_1|}\omega\tau_{r0})^{2d/(3d-2)}.
\label{n1}
\end{eqnarray}
Assuming $\sqrt{|\gamma_1|}\omega\tau_{r0}\gg 1$ we see that $n_L\gg
n_0$.

Eq.(\ref{nstar}) can be rewritten as
$n_{*}\!\approx\!n_0|\gamma_1|(\omega\tau_{r0})(\mu_0/\hbar\omega)$,
and the condition $|\gamma_1|\!\gg\!(\hbar\omega/\mu_0)^2$ (see
above)
ensures the inequality $n_L\!\ll\!n_{*}$ which justifies the use of
the
quasistationary approach in deriving Eqs.~(\ref{nmax1}), (\ref{t1})
and
(\ref{n1}).

The condition $n_L\ll n_{*}$ means that the collapse continues to
occur in the quasistationary regime also at times $\Delta t<\Delta
t_L$.
In this time interval, using Eqs.~(\ref{bstar}), (\ref{nmax}),
(\ref{t1}), and Eq.(\ref{aloss}) written in the form $\dot N\approx
-\alpha_{r}n^2(t)N(t)$, for the number of particles we find
\begin{equation}          \label{Nt}
N(t)\approx N_0(\Delta t/\Delta t_L)^{(3d-2)/4}.
\end{equation}
The condensate density increases as
\begin{equation}           \label{nt}
n(t)\approx n_0\sqrt{\tau_{r0}/\Delta t},
\end{equation}
i.e. slower than in the initial stage of the compression.

At times very close to $t_{*}$, i.e., at $\Delta t\approx\tau_{r*}
=(\alpha_rn_{*}^2)^{-1}$, the density approaches $n_{*}$, and the
picture of collapse changes.
If $n$ becomes larger than $n_{*}$, then the decrease of $n$ due to
particle losses already dominates over the dynamic compression.
On the other hand, for $n<n_{*}$ the dynamic compression dominates
and the
density rises.
Therefore, $n_{*}$ proves to be a critical density value which is
approximately
conserved in the final stage of collapse.
The decrease of the number of particles is determined by the
relation
$\dot N=-N/\tau_{r*}$, and we obtain
\begin{equation}             \label{Nf}
N(t)\approx N_{*}\exp{\{-(t-t_{*})/\tau_{r*}\}},
\end{equation}
where $N_{*}\ll N_0$ and is determined by Eqs.~(\ref{Nt}) and
(\ref{t1}),
with $\Delta t\approx\tau_{r*}$.
The characteristic size of the condensate decreases as
$N(t)/n_{*}$.

The above described picture of collapse is somewhat idealized.
For example, the mean interparticle separation at limiting density,
$n_{*}^{-1/3}$, can become
comparable with the characteristic radius of interatomic
interaction.
Nevertheless, we believe that the main results should remain
unchanged.
The scenario of collapse can be observed by measuring the
recombination
losses of particles.

Considering the global collapse in an anisotropic harmonic
potential we should return to scaling equations (\ref{bfast}), with
$\gamma_1<0$.
The rate of the self-compression is the highest in the direction
corresponding to the largest frequency.
In the case of cylindrical symmetry with the ratio of the axial to
radial
frequency, $\beta=\omega_z/\omega_{\rho}<1$, there will be a
characteristic
time $t_{*}$ at which $b_{\rho}(t_{*})=0$, but $b_z(t_{*})$ remaines
finite.
In the time interval, where $b_{\rho}(t)\ll 1$,
Eq.(\ref{bfast}) for $b_{\rho}(t)$ is reduced to that for the
isotropic
$2$-d collapse, with $\gamma_1$ replaced by $\gamma_1/b_z(t_{*})$.
Thus, the global collapse will occur along the lines of the above
described
$2$-d scenario, in combination with a ``slow'' compression in the
axial
direction.
For $\beta>1$ the condensate predominantly collapses in the axial
direction
and, due to a quasi-one-dimensional character of the
self-compression,
the kinetic energy term in Eq.(\ref{chigen}) increases. Eventually
the
parameter $\varepsilon(t)$ becomes of order unity,
and our scaling approach breaks down.
Such a rise of the kinetic energy prevents the further compression
and
can lead to a non-trivial oscillatory character of the global
collapse.

Density fluctuations on the background of
the global collapse can lead to the appearence of local collapses,
with
linear dimensions of order the healing length.
Although their influence on the global collapse can be reduced by
decreasing the ratio $n_0\tilde U/\hbar\omega$,
the study of the role of the local collapses requires a separate
analysis.

We acknowledge discussions with J.T.M. Walraven and M.W. Reynolds.
This work was supported by the Dutch Foundation FOM, by NWO
(project 047-003.036), by INTAS, and by the Russian Foundation for
Basic Studies.

\end{document}